\documentclass{PoS}
\usepackage{latexsym}
\usepackage{graphicx}
\usepackage{amsmath}%

\usepackage{amsfonts}%
\usepackage{amssymb}%

\PoS{PoS(JHW 2005)}

\newcommand{\beq}{\begin{equation}}
\newcommand{\eeq}{\end{equation}}
\newcommand{\ber}{\begin{eqnarray}}
\newcommand{\eer}{\end{eqnarray}}
\newcommand{\berr}{\begin{eqnarray*}}
\newcommand{\eerr}{\end{eqnarray*}}

\newcommand{\sg}{\sigma(\omega,\vec{p})}
\newcommand{\sgh}{\sigma_H(p_0,\vec{p})}

\FullConference{ 29th Johns Hopkins Workshop in Theoretical Physics\\
Budapest, Hungary\\
1-3 August 2005
}
\author{A.~Jakov\'ac\\
Institute of Physics, BME Budapest, Budafoki \'ut 8, H-1111 Budapest,
Hungary\\
\email{Antal.Jakovac@cern.ch}}
\author{P.~Petreczky\\
Physics Department, Brookhaven National Laboratory, Upton, NY, 11973,  USA\\
\email{petreczk@bnl.gov}}
\author{ \speaker{K.~Petrov},\\
Niels Bohr Institute, Copenhagen 2100, Denmark\\
E-mail: \email{kpetrov@nbi.dk}}
\author{A.~Velytsky\\
Department of Physics and Astronomy, UCLA, Los Angeles, CA 90095-1547, USA\\
\email{vel@physics.ucla.edu}}

\title{On Charmonia Survival Above Deconfinement}
\ShortTitle{On Charmonia Survival Above Deconfinement}

\abstract{
We study charmonium correlators and spectral functions at zero and finite temperature
using anisotropic lattices at several different lattice spacings. We find evidence for 
survival of 1S charmonia states at leas till $1.5T_c$ and dissolution of 1P states 
at $1.16T_c$.
 % end of PACS codes
} %end of abstract

\begin{document}

\section{Introduction}
\label{intro}

Heavy quarkonia can play an important role in studying hot and dense 
strongly interacting matter. Because of the heavy quark mass quarkonia binding can be understood
in terms of the static potential. Also becasue of their relatively small size they do not experience the
medium till relatively high temperatures. 
General considerations suggest  that quarkonia will melt at
temperatures right above the deconfinement temperature
as a result of modification of inter-quark forces (color screening).
Thus it was conjectured by Matsui and Satz \cite{MS86} that melting of
different quarkonia states due to color screening 
can signal Quark Gluon Plasma formation in heavy ion collisions.
Classic studies  of quarkonia dissolution 
\cite{karsch88,digal01a,digal01b,shuryak04,wong04,mocsyhard04} rely heavily on potential models. 
However it is very unclear if such models are valid 
at finite temperature  \cite{petreczkyhard04}. As an alternative method to approach this problem we may use 
lattice to determine meson spectral functions. For charmonium such studies
appeared only recently and suggested, contrary to potential models,
that $J/\psi$ and $\eta_c$ survive at temperatures as high as $1.6T_c$ 
\cite{umeda02,asakawa04,datta04}. It has been also found that $\chi_c$ 
melts at temperature of about  $1.1T_c$ \cite{datta04}. 
In  principle, meson spectral functions offer an attractive possibility to
study the properties of radially excited quarkonium states. In particular
it is straightforward to determine quarkonium decay width to dilepton.
In recent studies of charmonium spectral functions several peaks were
identified beyond and ground state \cite{umeda02,asakawa04,datta04}.
It is not clear, however, if
any of those peak correspond to radial excitation. Moreover it has
been argued that all peaks in the spectral functions except the first one
are lattice artifacts \cite{datta04}.

In this contribution we focus on our updated results for the charmonium; bottonium results and detailed study 
of the Maximum Enthropy Method reliability will be presented in our upcoming paper. Preliminary results for
bottomonium can be also found in Ref. \cite{lat05}.

\section{Meson correlators and spectral functions}%

In our lattice investigation we calculate correlators of point
meson operators of the form 
\begin{equation}
	J_H(t,x)=\bar q(t,x) \Gamma_H q(t,x),
\end{equation}
where $\Gamma_H=1,\gamma_5, \gamma_{\mu}, \gamma_5 \gamma_{\mu}, 
\gamma_{\mu} \gamma_{\nu}$ 
fixes the quantum number of the channel to 
scalar, pseudo-scalar, vector, axial-vector and tensor channels correspondingly.
The relation of these quantum number channels to different meson states is given
in Tab. \ref{tab.channels}.
\begin{table}[ht]
	\begin{center}
\begin{tabular}[c]{|c|c|c|c|}
	\hline
$\Gamma$ & $^{2S+1}L_{J}$ & $J^{PC}$ &\\\hline
$\gamma_{5}$ & $^{1}S_{0}$ & $0^{-+}$  \\
$\gamma_{s}$ & $^{3}S_{1}$ & $1^{--}$  \\
$\gamma_{s}\gamma_{s^{\prime}}$ & $^{1}P_{1}$ & $1^{+-}$  \\
$1$ & $^{3}P_{0}$ & $0^{++}$ \\
$\gamma_{5}\gamma_{s}$ & $^{3}P_{1}$ & $1^{++}$ \\
&&$2^{++}$\\\hline
\end{tabular}
\begin{tabular}
[c]{|cc|}\hline
$c\overline{c}(n=1)$ & $c\overline{c}(n=2)$\\\hline
$\eta_{c}$ & $\eta_{c}^{^{\prime}}$\\
$J/\psi$ & $\psi^{\prime}$\\
$h_{c}$ & \\
$\chi_{c0}$ & \\
$\chi_{c1}$ & \\
$\chi_{c2}$ & \\\hline
\end{tabular}

\caption{Charmonia states in different channels
\label{tab.channels}}
	\end{center}
\end{table}

Most dynamic properties of a finite temperature system are incorporated 
in the spectral function. The spectral function $\sgh$ for a given 
mesonic channel $H$ in a system at temperature $T$ can be defined 
through the Fourier transform of the real time two-point functions
$D^{>}$ and $D^{<}$ or, equivalently, as the imaginary part of 
the Fourier transformed retarded 
correlation function \cite{lebellac},
\ber
\sgh &=& \frac{1}{2 \pi} (D^{>}_H(p_0, \vec{p})-D^{<}_H(p_0, \vec{p}))=
\label{eq.defspect}\\
&&
\frac{1}{\pi} Im D^R_H(p_0, \vec{p}) \nonumber \\
 D^{>(<)}_H(p_0, \vec{p}) &=& \int{d^4 p \over (2
\pi)^4} e^{i p.x} D^{>(<)}_H(x_0,\vec{x}) \\
D^{>}_H(x_0,\vec{x}) &=& \langle
J_H(x_0, \vec{x}), J_H(0, \vec{0}) \rangle \nonumber\\
D^{<}_H(x_0,\vec{x}) &=& 
\langle J_H(0, \vec{0}), J_H(x_0,\vec{x}) \rangle , \quad x_0>0 \
\eer

The Euclidean time correlator calculated on the lattice
\beq
G_H(\tau, \vec{p}) = \int d^3x e^{i \vec{p}.\vec{x}} 
\langle T_{\tau} J_H(\tau, \vec{x}) J_H(0,
\vec{0}) \rangle
\eeq
is an analytic continuation of the real time correlator
$G_H(\tau,\vec{p})=D^{>}(-i\tau,\vec{p})$. 

Using this equation and the  Kubo-Martin-Schwinger
(KMS) condition \cite{lebellac} for the correlators
\beq
D^{>}_H(x_0,\vec{x})=D^{<}(x_0+i/T,\vec{x}),
\label{kms}
\eeq
one can relate the Euclidean propagator $G_H(\tau,\vec{p})$ to the 
spectral function, Eq. (\ref{eq.defspect}), through the integral
representation
\ber
G(\tau, \vec{p}) &=& \int_0^{\infty} d \omega
\sg K(\omega, \tau) \label{eq.spect} \nonumber\\
K(\omega, \tau) &=& \frac{\cosh(\omega(\tau-1/2
T))}{\sinh(\omega/2 T)}.
\label{eq.kernel}
\eer

To reconstruct the spectral function from the lattice correlator 
$G(\tau,T)$ this integral representation should be inverted. 
Since the number of data points is less than the number of degrees
of freedom (which is ${\cal O}(100)$ for reasonable discretization of
the integral ) spectral functions can be reconstructed only using the
Maximum Entropy Method (MEM) \cite{asakawa01}. In order to have sufficient
number of data points either very fine isotropic lattices \cite{datta04}
or anisotropic lattices \cite{umeda02,asakawa04} should be used.
Another difficulty arises due to the large quark mass;
discretization errors ${\cal O}(a m_{c,b})$ are present in the
heavy quark system. To remove these discretization errors we use
the Fermilab approach \cite{fermilab,chen01}.

We performed calculation of charmonia correlators in
quenched QCD using anisotropic lattices and Wilson gauge
action. For charmonium we use following gauge coupling values 
$\beta=5.6,~5.7,~5.9,~6.1$ with anisotropy $\xi=a_s/a_t=2$ and
$\beta=6.1,~6.5$ with $\xi=4$. When the lattice spacing is set using the Sommer scale
$r_0=0.5$ fm the above gauge coupling correspond to
lattice spacing $a_t^{-1}=1.56,~1.91,~2.91,~4.11,~8.18,~14.12$ GeV.
Further details of the lattice action together with the parameters 
can be found in Ref. \cite{jppv,manke}.

To be able to reconstruct the spectral functions
high statistical accuracy for charmonia correlators 
is required. This makes  
calculations computationally intensive, 
so we performed them on a prototypes of RBRC QCDOC machine.
It is a dedicated Lattice QCD machine developed by physicists from Columbia
University, BNL, RIKEN and UKQCD. Three such machines, each reaching about
10TFlops peak performance, are currently installed at BNL
and EPCC. Our prototypes were single-motherboard
machine at about 50 GFlops peak. Such resources are still not adequate for the
full QCD simulations, so we use quenched approximation, which is equivalent to
neglecting quark loops. Typical statistics gathered was 500 to 1000
measurements, separated by 400 updates.

\section{Bayesian analysis of meson correlators}
In Bayesian analysis of the correlator one looks for a spectral function which 
maximizes the 
conditional probability $P[\sigma|DH]$ of having the spectral function $\sigma$ given
the data $D$ and some prior knowledge $H$  (for a reviews see \cite{asakawa01,Lepage:2001ym}).
Different Bayesian  methods differ in the choice of the prior knowledge.
One version of this analysis which is extensively used in the literature is the 
{\em Maximum Entropy Method} (MEM) \cite{lat05,umeda02,asakawa04,datta04,asakawa01}.
In this method the basic prior knowledge is the positivity of the spectral function and 
the prior knowledge is given by the Shannon - Janes entropy  
\beq
S=\int d \omega \biggl [ \sigma(\omega)-m(\omega)-\sigma(\omega)
  \ln(\frac{\sigma(\omega)}{m(\omega)}) \biggr]. 
\eeq
The real function $m(\omega)$ is called the default model and parametrizes all additional prior knowledge, about the
spectral functions, such as the asymptotic behavior at high energy  \cite{asakawa01}.
For this case the conditional probability
\beq
 P[\sigma|DH]=\exp(-\frac{1}{2} \chi^2 + \alpha S),
\label{eq:PDH}
\eeq
with $\chi^2$ being the standard likelihood function and $\alpha$ a real parameter.
To maximize $P[\sigma|DH]$ usually the Bryan algorithm \cite{bryan} is used.
We use a new approach (to be described in \cite{jppv}) which through exact mathematical 
transforamtions reduces the problem to the task of minimization of a function with positive 
definite second derivative. This is a minimization problem in number-of-data dimensions,
which is not any more difficult task than applying a $\chi^2$ method. The Bryan algorithm
takes different approach of singular value decomposition in order to find a relevant subspace,
where minimization is performed.

\section{Spectral functions at zero temperature}
The spectral function for pseudo-scalar charmonium spectral functions is shown in Fig. 1.
The first peak in the spectral function corresponds to $\eta_c(1S)$ state.  The position of the
peak and the corresponding amplitude (i.e. the area unders the peak) are in good agreement
With the results of simple exponential fit. The second peak in the spectrak function is most likely
the combination of the excited states as its position and amplitude is higher than what one would
expect for 2S state. The spectral function becomes sensitive to the effects of the lattice spacings
for $\omega>5$GeV. In this $\omega$ region the spectral functions becomes sensitive to the choice
of the default model. This is because only a very few data points in the correlator carry information 
about the spectral function in that region. 

Also shown in Fig.1 is the spectral function in the
scalar channel. The 1st peak corresponds to $\chi_{c0}(1P)$ state. The correlator is more noisy
in the scalar channel than in the pseudo-scalar one. As the results the $\chi_{c0}(1P)$ peak
is less pronounced and has larger statistical errors. The peak position and the area unders the peak
is consistent with the simple exponential fit. As in the psudo-scalar case individula excited states are
not resolved and the spectral function depends on the lattice spacing and default model for $\omega>5$GeV. 

\begin{figure}
	\includegraphics[width=7cm]{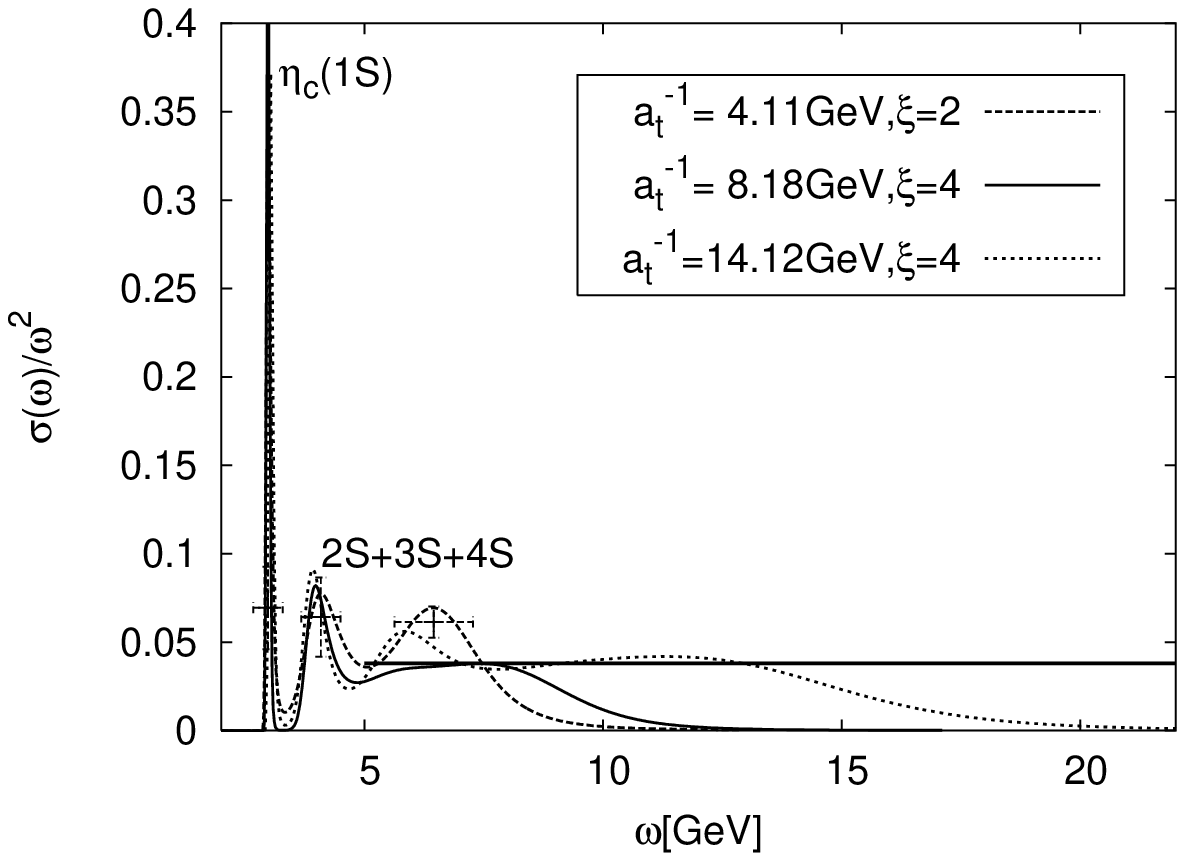}
	\includegraphics[width=7cm]{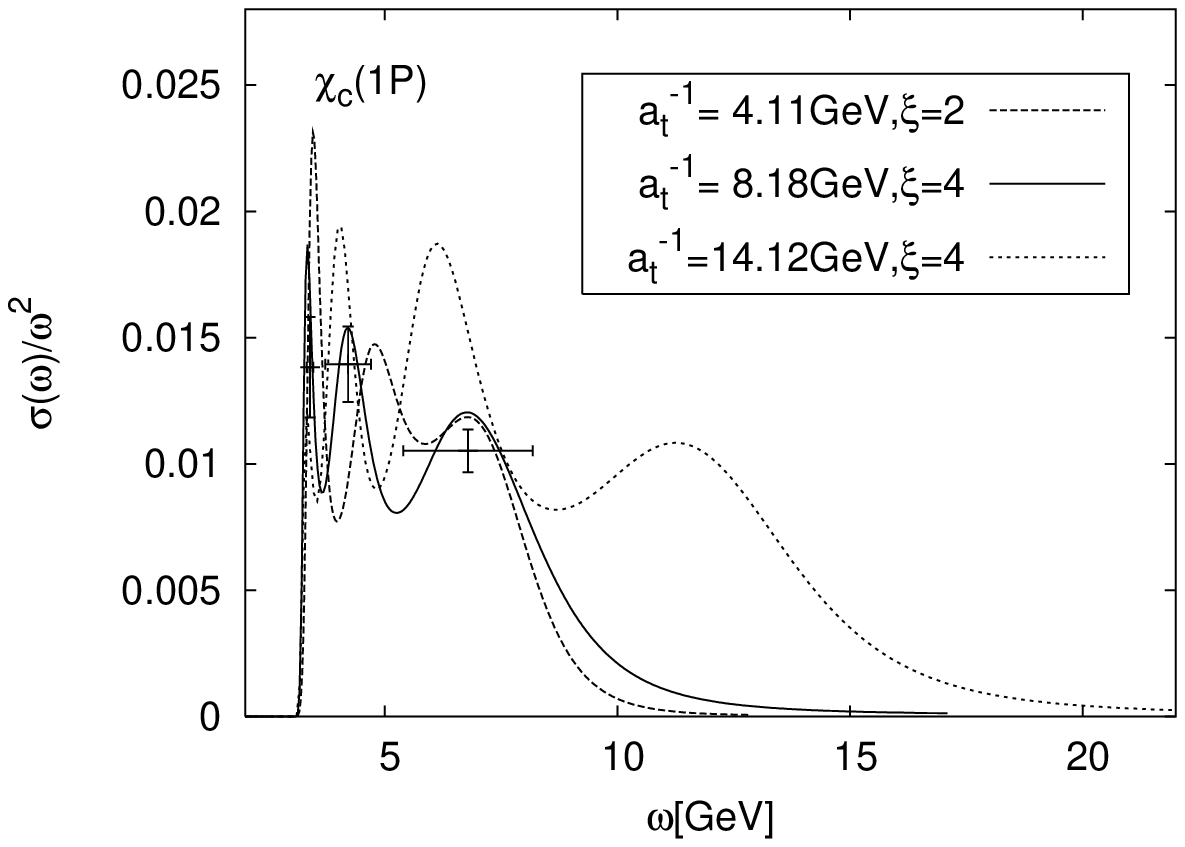}
\caption{
Charmonium spectral function in  the psuedoscalar channel (left) and 
the scalar channel (right) at different lattice spacings and zero temperature. }
\label{fig.spfct0}
\end{figure}

Similar results have been found for the vector and axial-vector
channels which correspond to $J/\psi$ and $\chi_{c1}$ states respectively.

We would like to know what happens to different charmonia states
at temperatures above the deconfinement temperature $T_c$. With
increasing temperature it becomes more and more difficult to
reconstruct the spectral functions as both the number of available
data points as well as the physical extent of the time direction
(which is $1/T$) decreases. Therefore it is useful to study the
temperature dependence of charmonia correlators first. From
Eq. (\ref{eq.kernel}) it is clear that the temperature dependence 
of charmonia correlators come from two sources: the temperature
dependence of the spectral function and temperature dependence of
the integration kernel $K(\tau,\omega,T)$. To separate out the 
trivial temperature dependence due to the integration kernel,
following Ref \cite{datta04} at each temperature we calculate
the so-called reconstructed correlator  
\begin{equation}
G_{recon}(\tau,T)=\int_{0}^{\infty}d\omega
\sigma(\omega,T=0)K(\tau,\omega,T)
\end{equation}
Now if we assume that there is no temperature dependence 
in the spectral function - then the ratio of the original and 
the reconstructed correlator should be close to one,
$G(\tau,T)/G_{recon}(\tau,T) \sim 1$. 
This way we can identify the cases when spectral function itself 
changes dramatically with temperature. 
This gives reliable information about the fate of charmonia states above
deconfinement. 

\begin{figure}
\includegraphics[width=3in]{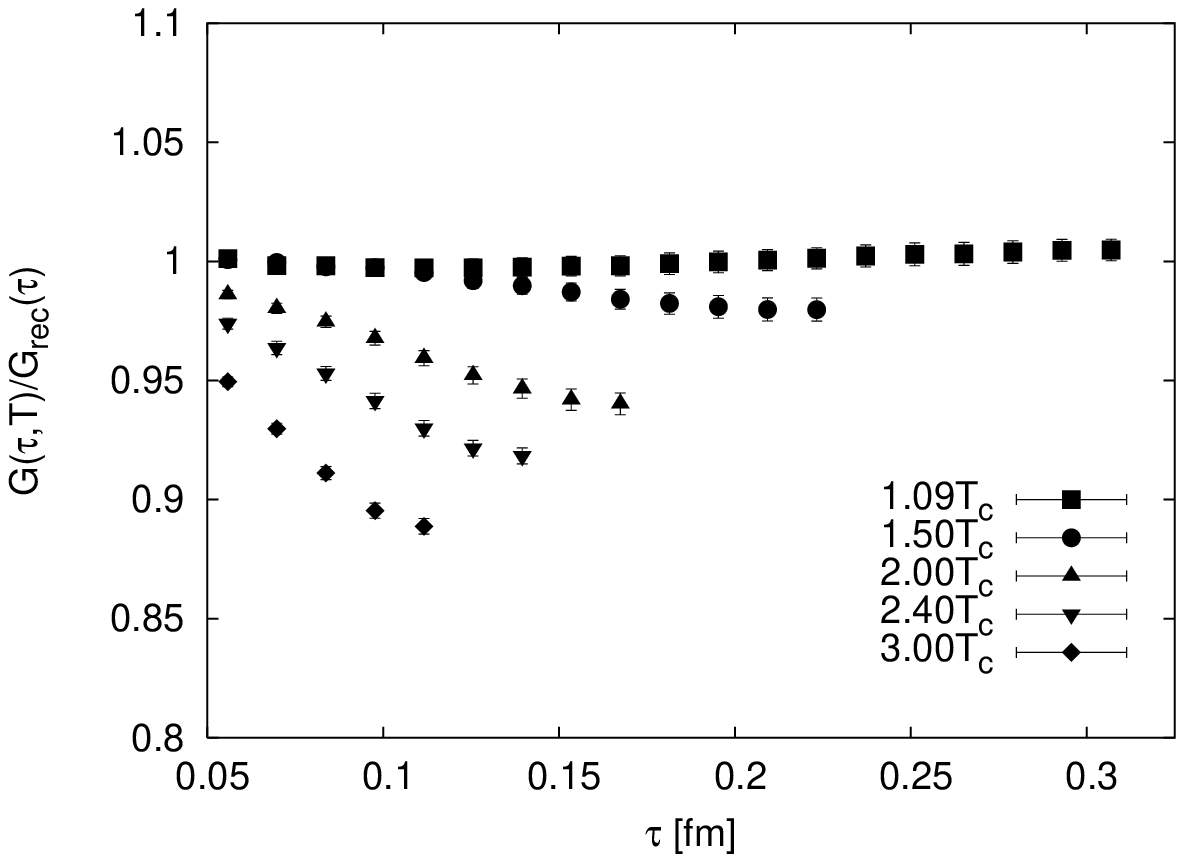}
\includegraphics[width=3in]{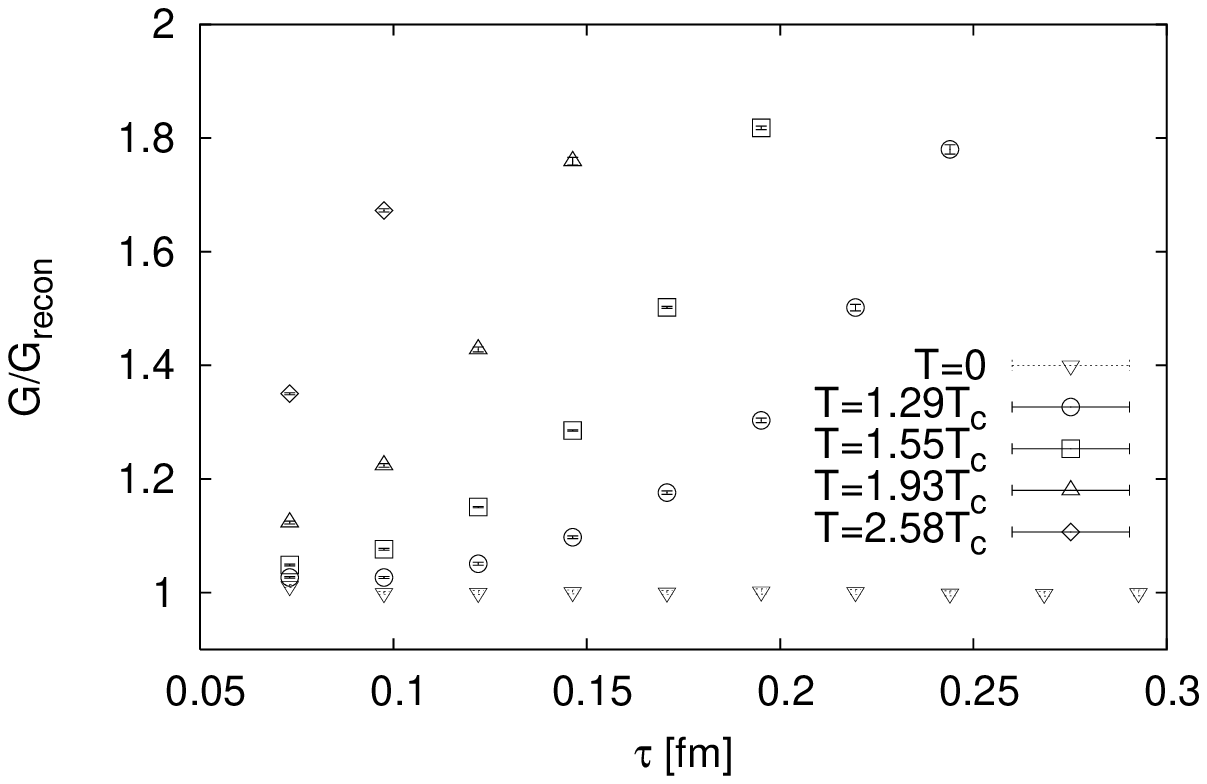}
\caption{The ratio $G(\tau,T)/G_{recon}(\tau,T)$ of charmonium for pseudoscalar channel at  at $a_t^{-2}=14.11$GeV
(left) and scalar channel at  at $a_t^{-2}=8.18$GeV  (right) at different
temperatures.}
\label{ratioc} 
\end{figure}

In Fig.~\ref{ratioc} we show this ratio at $\beta=6.5$ and $6.1$ for pseudoscalar and scalar 
channels of charmonium correspondingly. 

>From the figures one can see that the pseudo-scalar correlators shows
only very small changes till $1.5T_c$ indicating that the $\eta_c$ states survives
till this temperature with little modification of its properties. On the other hand the scalar
correlator shows large changes already at $1.16T_c$ suugesting strong modification or
dissolution of the  $\chi_{c0}$ state at this temperature.

\begin{figure}
	\includegraphics[width=7cm]{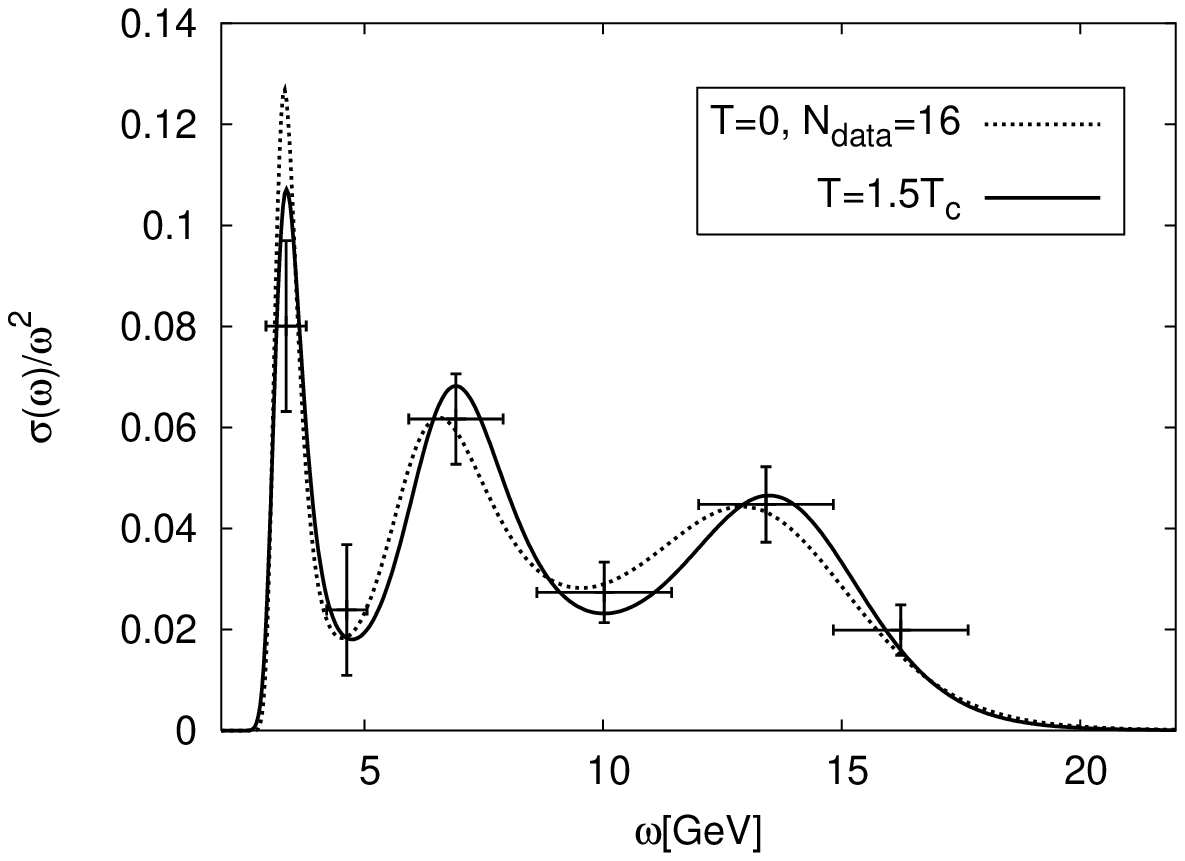}
	\includegraphics[width=7cm]{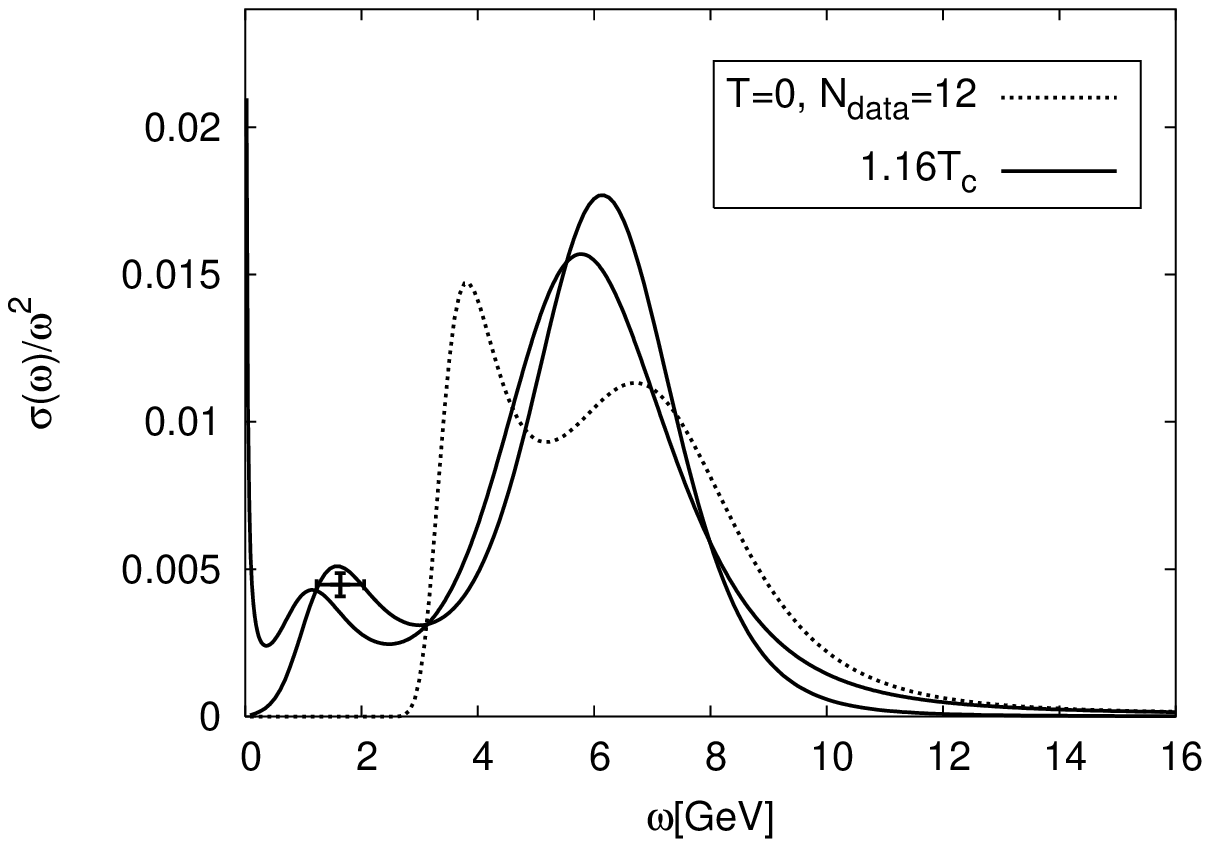}
\caption{
Charmonium spectral function in  the psuedoscalar channel at $a_t^{-2}=14.11$GeV  (left) and 
the scalar channel (right)   at $a_t^{-2}=8.18$GeV  at zero and above deconfinement temperatures. For finite temperature 
scalar chanel two different default  models are shown.}
\label{fig.spfct}
\end{figure}

More detailed information on different charmonia
states at finite temperature can be obtained by
calculating spectral functions using MEM.
The result of these calculation is show in
Figs.~\ref{fig.spfct}. 
Because at high temperature the temporal extent and the number of data
points where the correlators are calculated become smaller the spectral functuon reconstructed
using MEM are less reliable.  To study the temperature modifications of the spectral function
we compare the finite temperature spectral functions against the zero spectral functions
obatined from the correlator  using the same time interval and number of data points available
at finite temperature. 
We see that spectral function in the pseudo-scalar channel show no temperature dependence
within the statistical errors shown in the figure in accord with the analysis of the correlation functions.
Also the spectral functions show very little dependence on the default model. 

The scalar spectral function on the other hand shows large changes at
$1.16T_c$ which is consistent with correlator-based analysis. Also default model dependence
of the scalar correlator is large above the deconfinement transition (c.f. Fig. 3, right).

\section{Acknowledgments} 
This work was supported by U.S. Department of Energy under 
Contract No. DE-AC02-98CH10886 and by SciDAC project. 
A.V. was partially supported by NSF-PHY-0309362.
K.P. is supported by Marie Curie Excellence Grant under contract MEXT-CT-2004-013510.
A.J. is supported by Hungarian Science Fund OTKA (F043465).
Simulations performed using Columbia Physics System (CPS) with high-performance 
clover inverter written by P.~Boyle and other parts 
by RBC collaboration. 
Special thanks to C.~Jung for his generous help with CPS.

\end{document}